\documentclass{emulateapj}

\usepackage{amssymb}
\usepackage{graphicx}

\slugcomment{Accepted by ApJ, November 27th, 2007}

\begin{document}


\title{Blue Compact Dwarf Galaxies with Spitzer: The Infrared/Radio Properties}

\author{Yanling Wu\altaffilmark{1}, V. Charmandaris\altaffilmark{2,3}, J.~R. Houck\altaffilmark{1}, 
  J. Bernard-Salas\altaffilmark{1},V.  Lebouteiller\altaffilmark{1}, B.~R. Brandl\altaffilmark{4}, 
  D. Farrah\altaffilmark{1}}

\altaffiltext{1}{Astronomy Department, Cornell University, Ithaca, NY
  14853}

\altaffiltext{2}{University of Crete, Department of Physics, GR-71003,
Heraklion, Greece}

\altaffiltext{3}{IESL/Foundation for Research and Technology - Hellas,
  GR-71110, Heraklion, Greece and Chercheur Associ\'e, Observatoire de
  Paris, F-75014, Paris, France}

\altaffiltext{4}{Leiden Observatory, Leiden University, P.O. Box 9513,
2300 RA Leiden, The Netherlands}

\email{wyl@astro.cornell.edu, vassilis@physics.uoc.gr,jrh13@cornell.edu,
  jbs@isc.astro.cornell.edu,vianney@isc.astro.cornell.edu, 
  brandl@strw.leidenuniv.nl,duncan@isc.astro.cornell.edu}

\begin{abstract}
  We study the correlation between the radio, mid-infrared and
  far-infrared properties for a sample of 28 blue compact dwarf (BCD)
  and low metallicity star-forming galaxies observed by {\em Spitzer}.
  We find that these sources extend the same far-infrared to radio
  correlation typical of star forming late type galaxies to lower
  luminosities. In BCDs, the 24$\mu$m (or 22$\mu$m) mid-infrared to
  radio correlation is similar to starburst galaxies, though there is
  somewhat larger dispersion in their q$_{24}$ parameter compared to
  their q$_{\rm FIR}$. No strong correlations between the q parameter
  and galaxy metallicity or effective dust temperature have been
  detected, though there is a hint of decreasing q$_{24}$ at low
  metallicities. The two lowest metallicity dwarfs in our sample,
  IZw18 and SBS0335-052E, despite their similar chemical abundance,
  deviate from the average q$_{24}$ ratio in opposite manners,
  displaying an apparent radio excess and dust excess respectively.

\end{abstract}

\keywords{
  galaxies:starburst ---
  galaxies: dwarf ---
  infrared: galaxies ---
  radio continuum: galaxies
  }

\section{Introduction}

More than 35 years ago, a correlation between the 10$\mu$m
mid-infrared (mid-IR) and 1.4\,GHz luminosities in galactic nuclei was
first noticed by \citet{van71}. In 1983, the launch of the {\em
  Infrared Astronomical Satellite (IRAS)} brought a new era in
collecting large samples of data in the infrared, and sparked numerous
studies investigating the above mentioned correlation in star forming
galaxies. Both the radio and infrared emission are closely
related to star formation activities. The radio continuum is known to
originate from two processes: thermal free-free emission from ionized
gas in H{\sc ii} regions and non-thermal synchrotron radiation from
relativistic electrons that are accelerated in the supernovae remnants
(SNRs) \citep[see][ for a review]{Condon92}. The thermal emission
usually has a rather flat spectrum with
f$_{\nu}$\,$\propto$\,$\nu^{-0.1}$ while the non-thermal component
often has a much steeper spectral slope with
f$_{\nu}$\,$\propto$\,$\nu^{-0.8}$.  In normal galaxies the relative
contribution of the two components varies with frequency and at
1.4\,GHz the radio continuum is dominated (at nearly $\sim$90\%) by
the non-thermal component. The infrared emission is due to the thermal
re-radiation of starlight from dust surrounding H{\sc ii} regions
\citep{Harwit75}.  Based mostly on {\em IRAS} data, it has been shown
that the far-infrared (FIR, 40-120\,$\mu$m) to radio correlation holds
well over a remarkably wide range of star forming galaxies, spanning
several orders of magnitude in luminosities \citep{Helou85, Dejong85,
  Condon91,Yun01}.  The availability of deep observations of distant
galaxies by the {\em Infrared Space Observatory (ISO)} and recently by
{\em Spitzer} \citep{Werner04} showed that the correlation between the
IR and radio emission is not limited to the local universe but also
extends in galaxies at higher redshifts \citep{Garrett02,Appleton04}.
It was also revealed that in addition to the FIR, the mid-IR emission
at $\sim$24$\mu$m correlates with the 20cm radio continuum, though
with more scatter \citep{Elbaz02,Gruppioni03,Appleton04,Wu05}. A
number of very deep {\em Spitzer} mid-IR and FIR surveys can now probe
a population of galaxies with low infrared luminosities for which
ancillary data, including deep radio imaging, are becoming available
\citep[e.g.][]{Jannuzi99,Sanders07,Rosenberg06}. This is particularly
interesting since in the near future $\sim20\mu$m is the longest
wavelength which will likely be probed by the James Web Space
Telescope.  It is thus instructive to examine the mid-IR to radio
correlation in more detail in these low luminosity nearby systems, for
which little is known to date.

As ubiquitous as the FIR/radio correlation appears to be, there are a
few significant deviations from it.  Radio excess exists in
some instances, such as galaxies hosting an active galactic nucleus
(AGN). External magnetic field compression due to the interaction with
nearby galaxies could also produce extra emission in the radio
continuum \citep{Miller01}. Conversely, synchrotron deficiency has been
found in some nascent starburst galaxies studied by
\citet{Roussel03,Roussel06} which was attributed to the lack of time
for massive young stars to evolve into supernovae (SN) since these
galaxies are just at the onset of a starburst episode. On the other
hand, in the infrared, dust emission can be damped in an
optically-thin environment, because the ultraviolet (UV) and optical
light may not be fully reprocessed by the dust, as seen in low
luminosity dwarf galaxies.

Blue compact dwarf galaxies are a group of extra-galactic objects that
are characterized by their blue optical colors, small sizes
($\le$1\,kpc) and low luminosities (M$_B>$-18). These galaxies do not
display any AGN signature\footnote{A possible exception among our
  sources is CG0752 in which the high ionization lines
  [NeV]$\lambda$14.3/24.3$\mu$m are detected \citep[see][]{Hao07}.} and
have recent bursts of star formation in a relatively unevolved
chemical environment. As such they have been proposed as nearby
analogs of star formation in young galaxies in the early universe.  In
a metal poor environment, star forming regions are usually optically
thin and emit less in the infrared.  However, \citet{Devereux89}
suggested that in a low luminosity galaxy, the radio emission also
decreases, and probably faster than the infrared.  The deficiency in
both the non-thermal radio and FIR emission may counterbalance each
other and result in a similar FIR/radio ratio to the one observed in
normal spiral galaxies \citep{Klein91}.  This was examined by a study
of star formation rates (SFRs) in BCDs performed by \citet{Hopkins02},
in which the authors found an excellent agreement between the SFRs
estimated from 1.4\,GHz and 60\,$\mu$m luminosities.  As
\citet{Bell03} has pointed out though, the FIR/radio correlation is
almost linear, not because the IR and radio emission reflect the SFRs
correctly, but because in low luminosity galaxies they are both
underestimated by similar factors.  This is in agreement with
\citet{Helou93}, who found from their modelling work in disk galaxies
that the transparency of the disk was about the same to both
re-emission processes.  \citet{Hunt05a} in their analysis of the
spectral energy distributions (SEDs) of low metallicity BCDs, noticed
that these systems do not follow several of the usual correlations
between the mid-IR, FIR and radio emission and display a scatter of a
factor of $\sim$10.

The {\em Spitzer} Space telescope has enabled us to
study the infrared properties of a large sample of BCDs, probing the
lower end of the luminosity and metallicity range. In this paper, the
fifth in a series \citep{Houck04b,Wu06,Wu07a,Wu07b}, we examine their
mid-IR and FIR to radio correlation extending the work of
\citet{Hopkins02}. We describe the sample selection and the observational
data in \S 2. A detailed study of mid-IR and FIR/radio correlation, as
well as its dependence with other parameters, such as metallicity and
dust temperature are presented in \S 3.  We also discuss two extreme
cases, IZw18 and SBS0335-052 in \S 3. We summarize our conclusions
in \S 4.

\section{Observations}

As part of the IRS\footnote{The IRS was a collaborative venture
  between Cornell University and Ball Aerospace Corporation funded by
  NASA through the Jet Propulsion Laboratory and the Ames Research
  Center.} \citep{Houck04a} Guaranteed Time Observation (GTO) program
(PID: 85), we have compiled a sample of BCD candidates ($\sim$64)
selected from the Second Byurakan Survey (SBS), Bootes void galaxies
\citep{Kirshner81,Popescu00}, and other commonly studied BCDs. These
sources are known to have low metallicities ranging from
0.03\,Z$_\odot$ to 0.5\,Z$_\odot$\footnote{Here we adopt the new
  oxygen solar abundance of 12+log(O/H)=8.69 \citep{Prieto01}.
  \citet{Wu07b} have derived neon and sulfur abundances for a sample
  of BCDs using the infrared data, but these abundances are not
  available for all of the sources in this study.}. Their 22\,$\mu$m
fluxes have been published in \citet{Wu06}. We also include 10
galaxies from \citet{Engelbracht05} (PID: 59), which are mostly BCDs
and starburst galaxies, and span a larger metallicity range
(0.03\,Z$_\odot$$\sim$1.5\,Z$_\odot$)\footnote{The galaxies in PID 59
  partly overlap with the BCDs in PID 85. We have also excluded some
  galaxies which the authors derived their 24\,$\mu$m fluxes by doing
  a color correction to the {\em IRAS} 25\,$\mu$m fluxes because of
  the aperture difference between these two band filters.}. For all
galaxies with 22\,(24)\,$\mu$m detections, we searched the literature
as well as the public archives (NVSS and FIRST) for 1.4GHz radio
continuum data. We restirct our sample to sources with both mid-IR and
radio detections which results in a sample of 23 galaxies. Finally, we
also include 5 galaxies that have 22\,$\mu$m detections and 1.4\,GHz
upper limits published by Hopkins et al. (2002) for comparison between
ours and Hopkins' samples. Note that the sample was not selected based
on infrared properties, but merely on BCD-type objects and the
availability of both infrared and radio data.  As a result, our sample
is not complete, but the large number of sources that only became
detectable in the infrared with {\em Spitzer}, allows us to probe the
properties of low luminosity dwarf galaxies, and provide statistically
meaningful results. The observational information for this sample and
previously published data are presented in Table \ref{tab1}, which
includes the positions of the sources, their mid-IR, FIR and 1.4GHz
flux densities, as well as oxygen abundances of the ionized gas.

All sources in this study have mid-IR flux measurements either at
22$\mu$m with the IRS red peak-up camera and/or at 24\,$\mu$m with
MIPS \citep{Rieke04}. The photometric fluxes of these two bands differ
by less than 10\% for galaxies in the local universe.  This was
confirmed by using a suite of $\sim$100 spectra of nearby
galaxies from our IRS/GTO database and calculating their synthetic 22
and 24$\mu$m fluxes after convolving the spectra with the
corresponding filter response curves. For consistency, in our analysis
we use the MIPS 24\,$\mu$m measurements, and the 22\,$\mu$m values are
only used when the 24\,$\mu$m values are not available. For sources
above the IRS 22$\mu$m and MIPS 24$\mu$m saturation limits we use the
IRS low resolution spectrum to estimate a synthetic 24\,$\mu$m flux.
We also obtained far-infrared fluxes for our sample from the archival
{\em IRAS} 60 and 100\,$\mu$m data \citep{Moshir90, Sanders03}.

Most of the 1.4\,GHz radio continuum data are from the NRAO VLA Sky
Survey (NVSS) \citep{Condon98}, while one is from the Faint Images of
the Radio Sky at Twenty cm (FIRST) \citep{Becker95}, along with some
individual observations (references in Table \ref{tab1}). A total of
seven sources were too faint and were not included in the NVSS
catalogue. For those we used the values of \citet{Hopkins02} who
studied a similar BCD sample and remeasured the 1.4GHz fluxes using
the images from NVSS and FIRST, providing radio detections for 2
sources and better upper limits for another five sources that overlaps
with the galaxies in our sample.

\begin{deluxetable*}{llrrrrrrccccc}
  \tabletypesize{\scriptsize}
  \setlength{\tabcolsep}{0.02in}
  \tablecaption{Properties of the Sample\label{tab1}}
  \tablewidth{0pc}
  \tablehead{
    \colhead{ID} & \colhead{Object Name} & \colhead{RA} & \colhead{Dec} & \multicolumn{4}{c}{Flux (mJy)} & 
    \colhead{12+log(O/H)} & \multicolumn{4}{c}{References} \\
    \cline{5-8}\cline{10-13}
    \colhead{} & \colhead{} & \colhead{J2000} & \colhead{J2000} & \colhead{1.4\,GHz} & \colhead{24(22)\,$\mu$m} & 
    \colhead{60\,$\mu$m} & \colhead{100\,$\mu$m} & \colhead{} & \colhead{1.4\,GHz} & \colhead{24\,$\mu$m} &
    \colhead{IRAS} & \colhead{Z} \\
  }
  \startdata
  
1  &  Haro11              & 00h36m52.5s  & -33d33m19s &  26.8  &   1900   &   6880 &   5040 &  7.9  & (1) & (2) & (3) & (4) \\
2  &  NGC1140             & 02h54m33.6s  & -10d01m40s &  23.6  &    316.6 &   3358 &   4922 &  8.5  & (1) & (5) & (6) & (7) \\
3  &  SBS0335-052E        & 03h37m44.0s  & -05d02m40s &  0.46  &     66   & \nodata& \nodata&  7.3  & (8) & (2) &     & (9) \\
4  &  NGC1569             & 04h30m47.0s  & +64d50m59s & 336.3  &   2991   &  54360 &  55290 &  8.2  & (1) & (2) & (3) &(10) \\
5  &  IIZw40              & 05h55m42.6s  & +03d23m32s &  32.5  &   1500   &   6570 &   5270 &  8.1  & (1) & (2) & (3) & (7) \\
6  &  UGC4274             & 08h13m13.0s  & +45d59m39s &  10.5  &    240   &   3236 &   6448 &  8.5  & (1) & (2) & (6) &     \\ 
7  &  He2-10              & 08h36m15.2s  & -26d24m34s &  83.8  &   4900   & \nodata& \nodata&  8.9  & (1) & (2) &     &(11) \\
8  &  NGC2782             & 09h14m05.1s  & +40d06m49s & 124.5  &    960   &   9170 &  13760 &  8.8  & (1) & (2) & (3) &(12) \\
9  &  NGC2903             & 09h32m10.1s  & +21d30m03s & 444.5  &   2200   &  60540 & 130430 &  9.3  & (1) & (2) & (3) &(11) \\
10 &  IZw18               & 09h34m02.0s  & +55d14m28s &   1.83 &      5.5 & \nodata& \nodata&  7.2  &(13) & (2) &     & (9) \\
11 &  SBS0940+544         & 09h44m16.7s  & +54d11m33s &$<$2.3  &      2.3 & \nodata& \nodata&  7.5  &(14) & (5) &     &(15) \\ 
12 &  NGC3077             & 10h03m19.1s  & +68d44m02s &  29.0  &   1500   &  15900 &  26530 &  8.6  & (1) & (2) & (3) &(16) \\
13 &  Mrk153              & 10h49m05.0s  & +52d20m08s &   4.0  &     29   &    280 & $<$480 &  7.8  & (1) & (2) & (6) &(17) \\
14 &  VIIZw403            & 11h27m59.9s  & +78d59m39s &   1.2  &     28   & \nodata& \nodata&  7.7  &(18) & (2) &     & (9) \\
15 &  Mrk1450             & 11h38m35.6s  & +57d52m27s &$<$2.0  &     48   &    279 & $<$575 &  8.0  &(14) & (2) & (6) & (7) \\
16 &  UM448               & 11h42m12.4s  & +00d20m03s &  32.6  &    560   &   4139 &   4321 &  8.0  & (1) & (2) & (6) & (7) \\
17 &  UM461               & 11h51m33.3s  & -02d22m22s &$<$2.6  &     30   & \nodata& \nodata&  7.8  &(14) & (2) &     & (9) \\
18 &  UM462               & 11h52m37.3s  & -02d28m10s &   5.8  &    110   &    944 &    896 &  8.0  & (1) & (2) & (6) & (9) \\
19 &  SBS1159+545         & 12h02m02.4s  & +54d15m50s &$<$2.3  &      6.4 & \nodata& \nodata&  7.5  &(14) & (5) &     &(19) \\ 
20 &  NGC4194             & 12h14m09.5s  & +54d31m37s & 100.7  &   3100   &  23200 &  25160 &  8.8  & (1) & (2) & (3) &(16) \\
21 &  NGC4670             & 12h45m17.1s  & +27d07m32s &  13.7  &    210   &   2634 &   4470 &  8.2  & (1) & (2) & (6) &(12) \\
22 &  SBS1415+437         & 14h17m01.4s  & +43d30m05s &   4.3  &     19.6 & \nodata& \nodata&  7.6  &(14) & (5) &     & (9) \\
23 &  Mrk475              & 14h39m05.4s  & +36d48m21s &$<$2.7  &     10.8 & \nodata& \nodata&  7.9  &(14) & (5) &     & (9) \\
24 &  CG0563              & 14h52m05.7s  & +38d10m59s &   6.2  &     97.0 &    870 &   1900 &  8.7  & (1) &     & (6) &     \\
25 &  CG0752              & 15h31m21.3s  & +47d01m24s &   4.8  &    138.9 &    837 &   1059 &\nodata& (1) & (5) & (6) &     \\
26 &  SBS1533+574         & 15h34m13.8s  & +57d17m06s &   4.2  &     53.4 &    257 &    405 &  8.1  &(14) & (5) & (6) & (9) \\
27 &  Mrk1499             & 16h35m21.1s  & +52d12m53s &   1.5  &     33.8 &    256 &    617 &  8.1  &(20) & (5) & (6) &(21) \\
28 &  Mrk930              & 23h31m58.3s  & +28d56m50s &  12.2  &    170   &   1245 &$<$2154 &  8.1  & (1) & (2) & (6) & (7) \\

\enddata

\tablerefs{(1) \citet{Condon98}, (2) \citet{Engelbracht05}, (3) \citet{Sanders03}, (4) \citet{Bergvall02}, 
  (5) \citet{Wu06}, (6) \citet{Moshir90}, (7) \citet{Guseva00}, (8) \citet{Hunt04}, (9) \citet{Izotov99},
  (10) \citet{Kobulnicky97}, (11) \citet{Kobulnicky99}, (12) \citet{Heckman98}, (13) \citet{Hunt05a}, 
  (14) \citet{Hopkins02}, (15) \citet{Thuan05}, (16) \citet{Storchi94}, (17) \citet{Kunth85}, (18) \citet{Leroy05}, 
  (19) \citet{Izotov98}, (20) FIRST catalog, (21) \citet{Shi05}}
\end{deluxetable*}

Our final sample consists of 28 galaxies, all of which have {\em Spitzer}
mid-IR 24\,$\mu$m and/or 22\,$\mu$m flux measurements.  Among these
galaxies, 23 sources have 1.4GHz radio continuum data and 5 have
measured upper limits. {\em IRAS} 60\,$\mu$m and 100\,$\mu$m fluxes
are available for 16 sources and 3 more are detected only at
60\,$\mu$m. We list the photometry of our sources in Table \ref{tab1}.
The uncertainty in the 22\,$\mu$m or 24\,$\mu$m photometry is less
than 5\%.  The {\em IRAS} 60\,$\mu$m and 100\,$\mu$m fluxes typically
have less than a few percent of error, but could go up to $\sim$15\%
for some of the fainter sources.  The rms noise level for NVSS is
$\sim$0.5\,mJy\,beam$^{-1}$.

\section{Results}

\subsection{Mid-IR and FIR to Radio Correlation in BCDs}

The sensitivity and efficiency of the {\em Spitzer} Space Telescope
has allowed us to probe the correlation between the mid-IR and radio
luminosities for a large number of galaxies. Using the 24\,$\mu$m and
70\,$\mu$m MIPS imaging data of the First Look Survey,
\citet{Appleton04} have demonstrated the first direct evidence for the
universality of the mid-IR/radio and FIR/radio correlation to
z$\sim$1. \citet{Wu05} have also studied the mid-IR/radio correlation
in a sample of star-forming galaxies and found that both the 8\,$\mu$m
and 24\,$\mu$m luminosity are clearly correlated with the 1.4GHz radio
luminosity. Their sample included only a few (3) dwarf galaxies and
suggested that there may be a slope change for dwarf galaxies,
which could be due to the lower dust-to-gas ratios and lower metallicities
of the dwarfs.  A detailed analysis of the spatial distribution of the
infrared to radio correlation using {\em Spitzer} data on a sample of nearby
late type spiral galaxies has been performed by
\citet{Murphy06a,Murphy06b} in which the authors found that the ratio
of the mid-IR and FIR to radio emission does vary from one region to
the other. In general though, the dispersion is small and the ratio
decreases with surface brightness and galactocentric radius.

\begin{figure}
  \epsscale{1.2}
  \plotone{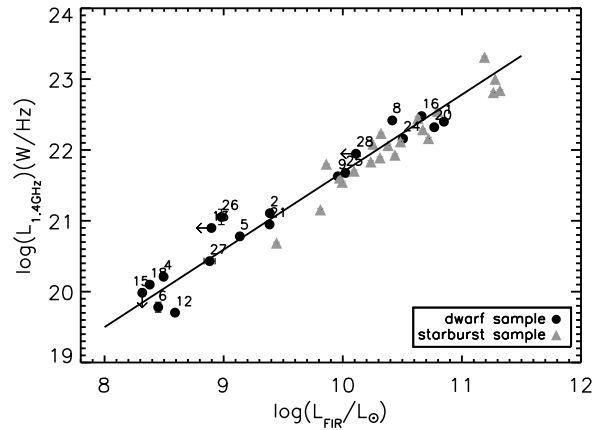}
  \caption{The FIR to 1.4\,GHz radio luminosity correlation for this
    sample. Our data are presented with filled circles. The numbers
    next to each symbol correspond to their IDs in Table \ref{tab1}.
    The solid line is the best fit to this sample excluding the upper
    limits.  For comparison, we have also include the starburst
    galaxies of \citet{Brandl06}, indicated with triangles.}
  \label{fig:fig1}
\end{figure}

\begin{deluxetable*}{llrcccrr}
  \tabletypesize{\scriptsize}
  \setlength{\tabcolsep}{0.02in}
  \tablecaption{Derived Quantities of the Sample\label{tab2}}
  \tablewidth{0pc}
  \tablehead{
    \colhead{ID} & \colhead{Object Name} & \colhead{Distance\tablenotemark{a}} & \colhead{L$_{\rm 1.4GHz}$} 
    & \colhead{L$_{24\mu m}$} & \colhead{L$_{\rm FIR}$} & \colhead{q$_{24}$} & \colhead{q$_{\rm FIR}$} \\
    \colhead{} & \colhead{} & \colhead{(Mpc)} & \colhead{($\times$10$^{20}$\,W\,Hz$^{-1}$)} &
    \colhead{($\times$10$^8$L$_\odot$)} & \colhead{($\times$10$^8$L$_\odot$)} & \colhead{} & \colhead{}
  }
  \startdata
  
1  &  Haro11              &  88   & 250.5  & 584.2  &   706.3  & 1.85  & 2.46  \\
2  &  NGC1140             &  21.2 & 12.8   & 5.6    &   24.4   & 1.13  & 2.29  \\
3  &  SBS0335-052E        &  58   & 18.7   & 8.8    &  \nodata & 2.16  &\nodata\\
4  &  NGC1569             &  2.0  & 1.6    & 0.5    &   3.1    & 0.95  & 2.29  \\
5  &  IIZw40              &  12.4 & 6.0    & 9.2    &   13.7   & 1.66  & 2.36  \\
6  &  UGC4274             &  6.9  & 0.6    & 0.5    &   2.8    & 1.36  & 2.68  \\ 
7  &  He2-10              &  12   & 15.6   & 29.9   &  \nodata & 1.77  &\nodata\\
8  &  NGC2782             &  41.7 & 216.3  & 66.3   &   260.4  & 0.89  & 2.00  \\
9  &  NGC2903             &  8.9  & 42.5   & 6.9    &   90.9   & 0.69  & 2.34  \\
10 &  IZw18               &  18.2 & 0.7    & 0.07   &  \nodata & 0.48  & 1.31\tablenotemark{b}\\
11 &  SBS0940+544         &  23   &$<$ 1.5 & 0.05   &  \nodata &$>$0.00&\nodata\\ 
12 &  NGC3077             &  3.8  & 0.5    & 0.9    &   3.9    & 1.71  & 2.89  \\
13 &  Mrk153              &  40.5 & 7.9    & 1.9    &  $<$7.9  & 0.86  &$<$2.00\\
14 &  VIIZw403            &  4.3  & 0.03   & 0.02   &  \nodata & 1.37  &\nodata\\
15 &  Mrk1450             &  20.0 &$<$1.0  & 0.8    &  $<$2.1  &$>$1.38&\nodata \\
16 &  UM448               &  87.4 & 300.6  & 188.9  &   458.6  & 1.23  & 2.19  \\
17 &  UM461               &  13.4 &$<$0.6  & 0.2    &  \nodata &$>$1.06&\nodata\\
18 &  UM462               &  13.4 & 1.3    & 0.8    &   2.4    & 1.28  & 2.29  \\
19 &  SBS1159+545         &  50   &$<$7.0  & 0.6    &  \nodata &$>$0.44&\nodata\\ 
20 &  NGC4194             &  41.5 & 209.3  & 212.0  &   586.0  & 1.49  & 2.45  \\
21 &  NGC4670             &  23.2 & 8.9    & 4.5    &   24.3   & 1.19  & 2.44  \\
22 &  SBS1415+437         &  8.7  & 0.4    & 0.06   &  \nodata & 0.66  &\nodata\\
23 &  Mrk475              &  11.2 &$<$0.4  & 0.05   &  \nodata &$>$0.60&\nodata\\
24 &  CG0563              &  139  & 144.6  & 74.4   &   320.5  & 1.19  & 2.35  \\
25 &  CG0752              &  90   & 47.3   & 45.1   &  105.3   & 1.46  & 2.35  \\
26 &  SBS1533+574         &  47   & 11.3   & 4.7    &    9.6   & 1.10  & 1.93  \\
27 &  Mrk1499             &  39   & 2.7    & 2.0    &   7.6    & 1.35  & 2.46  \\
28 &  Mrk930              &  77.5 & 88.4   & 40.5   &  $<$129.0& 1.14  &$<$2.17\\

\enddata

\tablenotetext{a}{The distances to the galaxies of the sample are
  adopted from \citet{Moustakas06} when available, while the rest of
  the sources are calculated from the redshifts taken from NED,
  assuming a $\Lambda$CDM cosmology with
  H$_0$\,=\,70~km~s$^{-1}$\,Mpc$^{-1}$, $\Omega_m$\,=\,0.3 and
  $\Omega_{\lambda}\,=\,0.7$. For IZw18, we adopt the newly derived
  distance by \citet{Aloisi07}. The average uncertainty in the
    distances is $\sim$5\%, mainly due to the value of H$_0$.}
\tablenotetext{b}{This is calculated based on the ``equivalent'' {\em
    IRAS} 60 and 100\,$\mu$m fluxes (see Section 3.4).}

\end{deluxetable*}

Using the data from this sample of low metallicity dwarf galaxies we
plot in Fig. 1 the radio luminosity of the sample as a function of the
FIR luminosity.  The luminosities of the sources we study span nearly
4 orders of magnitudes, but the correlation between the FIR and the
radio is remarkably tight.  The scatter in the ratio of the FIR and
radio luminosities is 0.23 dex, i.e. less than a factor of 2. We
performed a least-squares bisector fit to the data and found:
log[L$_{\rm 1.4GHz}$(WHz$^{-1}$)]=1.09$\times$log(L$_{\rm
  FIR}$/L$_\odot$)+10.75 .  For comparison, we have included in the
plot the starburst galaxies from \citet{Brandl06} marked with
triangles, which has an identical slope (within 1$\sigma$).  This
slope of 1.09$\pm$0.07 for the dwarf galaxy data agrees well with the
slope of 1.10$\pm$0.04 found by \citet{Bell03} for a sample of 162
galaxies, as well as the slope of 1.11$\pm$0.02 for the infrared
selected sources from the {\em IRAS} Bright Galaxy Sample (BGS)
\citep{Condon91}.  This is also in agreement with \citet{Hopkins02}
and would suggest that globally our BCDs have a very similar FIR/radio
correlation to normal galaxies.

Another way to parameterize the IR/radio correlation is to calculate
the ratio of FIR to radio luminosity $q_{\rm FIR}$ according to the
\citet{Helou85} formula as well as the $q_{24}$ following the
definition of \citet{Appleton04}\footnote{We use q$_{\rm
    FIR}$=log[1.26$\times 10^{-14}$ (2.58S$_{60\mu m}$+S$_{100\mu
    m}$)/(3.75 $\times 10^{12}$ F$_{1.4 \rm GHz})$] where S$_{60\mu
    m}$ and S$_{100\mu m}$ are in Jy and F$_{1.4 \rm GHz}$ is in
  Wm$^{-2}$Hz$^{-1}$ \citep{Helou85}. We also define $q_{24}$= log
  (S$_{24 \mu m}$/S$_{1.4 \rm GHz}$) where S$_{24 \mu m}$ and S$_{1.4
    \rm GHz}$ are in Jy as in \citet{Appleton04}.}. We plot the
$q_{24}$ values for our sample as a function of the 24$\mu$m
luminosity of the galaxies in Fig. 2. For this sample we find that
$q_{\rm FIR}=$ 2.4$\pm$0.2, consistent with the value of normal
galaxies of $q_{\rm FIR}=$ 2.3$\pm$0.2 found by \citet{Condon92}. When
using the mid-IR 24$\mu$m data, we find that $q_{24}=$1.3$\pm$0.4 (see
Table \ref{tab2}). The standard deviation in $q_{24}$ is $\sim$twice
that of $q_{\rm FIR}$. This is not unexpected given that the spectrum
of star forming galaxies shows substantially larger variations in
spectral slope in the mid-IR \citep[see][]{Brandl06} than in the FIR
\citep{Dale06}. A small change in the geometry of the emitting regions
would affect F$_{\nu}$(24\,$\mu$m) and thus q$_{24}$ ratio much more
than the FIR emission and q$_{\rm FIR}$.  A similar result has also
been noticed by \citet{Appleton04} and \citet{Murphy06a} who found a
larger dispersion in q$_{24}$ as compared to q$_{70}$\footnote{We
  define q$_{70}$= log (S$_{70 \mu m}$/S$_{1.4 \rm GHz}$) where S$_{70
    \mu m}$ and S$_{1.4 \rm GHz}$ are in Jy as in \citet{Appleton04}.}
and suggested that this is probably due to a larger intrinsic
dispersion in the IR/radio correlation at shorter wavelengths.
Interestingly, the \citet{Appleton04} values for q$_{24}=$
0.84$\pm$0.28 or the k-corrected q$_{24}$ of 0.94$\pm$0.23 are
somewhat smaller than our results, though consistent within 2$\sigma$.
One possible explanation is that the dust temperature of low
luminosity dwarf galaxies tends to peak at shorter wavelength than
normal star forming galaxies, which would result in an elevated
24\,$\mu$m based luminosity. An alternative explanation is that due to
the peculiar morphology of dwarf galaxies, more electrons escape from
the galaxy because of the cosmic ray diffusion.  \citet{Boyle07} have
also found a high q$_{24}$=1.39$\pm$0.02 for the sources they studied,
and postulated that this may be due to a change in the mean q$_{24}$
ratio for objects with F$_\nu$(24\,$\mu$m) $<$1\,mJy, however, their
sources are much further away and may not be comparable to the low
metallicity star forming galaxies we study in the local universe.

\begin{figure}
  \epsscale{1.2}
  \plotone{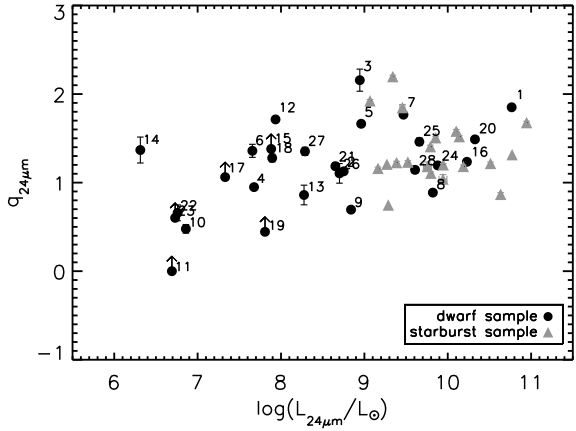}
  \caption{The $q_{24}$ as a function of the the 24\,$\mu$m
    monochromatic luminosity L$_{24}$. The symbols are
    the same as in Fig. 1.}
  \label{fig:fig2}
\end{figure}

\subsection{Metallicity and Dust Temperature Effects on q$_{IR}$}

A number of physical parameters, such as the metallicity, dust grain
size distribution, and temperature may affect the shape of the
infrared SED. Consequently, we search for correlations between the q
ratios of the galaxies in our sample with those parameters.  Our
sample covers a metallicity range of 7.2$\le$12+log(O/H)$\le$8.9. In
low metallicity galaxies, the dust content is usually low, even though
there are notable exceptions such as SBS0335-052E \citep{Houck04b},
thus the emission of UV light might not be fully reprocessed by the
dust and re-emitted at the infrared wavelengths.  However, these
galaxies might have a quenched synchrotron radiation. This can be
attributed to various reasons including a lack of supernovae remnants
which accelate particles producing radio emission, the escape of fast
cosmic rays from the galaxy, etc.  These two competing factors, dust
and radio emission, counter balance each other \citep[see][]{Bell03}.
BCDs though, are typically small, less than $\sim$1\,kpc in size, and
have fewer HII regions with massive stars (w.r.t. normal galaxies)
which heat the dust and go supernovae to produce cosmic rays. As a
result the averaging effects in the sampling of the properties of the
interstellar medium which result from the limited spatial resolution
in dense galactic disks is not so prominent in BCDs. Furthermore, the
small number of statistics may also contribute to the observed higher
dispersion in q$_{24}$ ratios in metal-poor dwarf galaxies.  We
investigate how metallicity affects the q$_{24}$ ratios by dividing
the dwarf galaxies in our sample into two groups: a lower metallicity
group with 12+log(O/H)$\le$8.0 and a higher metallicity group with
12+log(O/H)$>$8.0. This metallicity threshold was selected following
\citet{Rosenberg07} who noticed a change in the properties of the
star-forming dwarf galaxies they studied around this metallicity
value. We find a mean q$_{24}= $1.1$\pm$0.1\footnote{Here we use the
  standard error of the mean (SEM) to quantify the dispersion in the
  mean value.} for the first group and q$_{24}=$ 1.3$\pm$0.1 for the
second one.
If we calculate the q$_{24}$ for the low
metallicity group without including SBS0335-052E, we find a much lower
q$_{24}=$ 0.9$\pm$0.1. 
We also notice that it appears q$_{24}$ generally decreases with
reduced metallicity for sources with 12+log(O/H)$<$8.0, with
SBS035-052E as a clear outlier, and the slope flattens out at
12+log(O/H)$>$8.0 (see Fig. 3). This scatter in q$_{24}$ at higher
metallicities could be attributed to the more dispersion in the mid-IR
emission in these sources and is consistent with a number of other
studies that have found increased dispersion in the q ratios for very
high luminosity galaxies (which usually have higher metallicities)
\citep{Condon91, Yun01,Bell03}.

\begin{figure}
  \epsscale{1.2}
  \plotone{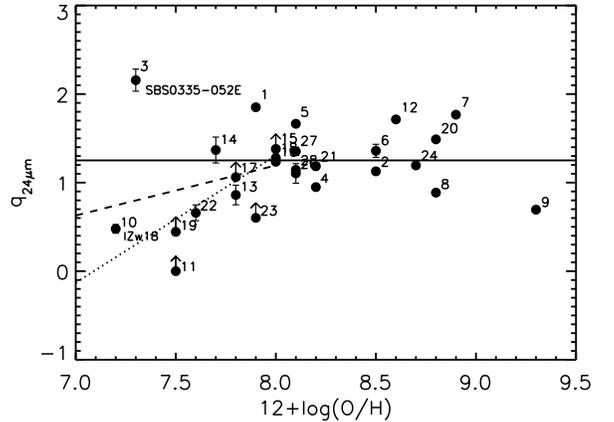}
  \caption{The q$_{24}$ ratio plotted as a function of the oxygen
    abundance of the BCDs. The mean q$_{24}$ value for all the sources
    is indicated by the solid line. A fit to the low metallicity
    sources (12+log(O/H)$\le$8.0) is indicated by the dashed line;
    while the dotted line is the same fit excluding SBS0335-052E. }
  \label{fig:fig3}
\end{figure}

\citet{Hummel88} have found evidence that, for a given galaxy, there
is a small decrease in F(100\,$\mu$m)/S(20\,cm) when the dust
temperature increases.  The latter can be traced by the ratio of
F(60\,$\mu$m)/F(100\,$\mu$m). \citet{Roussel03} reached a similar
result for the starburst galaxies they studied, in which they detected
an anti-correlation of q$_{\rm FIR}$ with
F(60\,$\mu$m)/F(100\,$\mu$m).  However, performing the same test for
the q$_{24}$ ratios on our dwarf galaxy sample which spans a
range of flux ratios -0.4 $<$log({\rm
  F(60\,$\mu$m)/F(100\,$\mu$m)})$<$0.2 we find no such clear
trend. It is conceivable that this is partly because our sample is too
small to identify a trend due to the intrinsic scatter in q$_{24}$.

\subsection{Star Formation Rate Estimates}

Both the radio and infrared emission can be used to estimate the star
formation rates (SFRs) in galaxies.  However, these correlations depend
on a number of parameters including dust content, optical depth and
metallicity. This topic and potential caveats have been discussed
extensively in the literature \citep[see][ and references
therein]{Condon92,Kennicutt98}.  For dwarf galaxies, the topic has
been addressed in detail by \citet{Hopkins02}. The recent wealth of
data from {\em Spitzer} has also provided sufficient motivation to
establish a calibration for the SFR using the infrared.  \citet{Wu05}
and more recently \citet{Calzetti07} have explored this topic using a
large sample of star forming regions and nearby galaxies.

In Fig. 4 we plot the IR estimated SFR for our sample using the
calibrations proposed by \citet{Wu05} and
\citet{Calzetti07} respectively:
\begin{equation}
{\rm SFR_{24}}(M_{\odot}yr^{-1})=\frac{\nu L_\nu(24 \mu m)}{6.66\times10^8L_\odot}
\end{equation}

\begin{equation}
{\rm SFR_{24}}(M_{\odot}yr^{-1})=1.27\times 10^{-38}[L_{24}(erg s^{-1})]^{0.885}
\end{equation}

as a function of the well-known radio to SFR formula of
\citet{Condon92}:

\begin{equation}
{\rm SFR_{1.4GHz}}=5.5\times\frac{L_{\rm 1.4GHz}}{4.6\times10^{21}(\rm{W Hz^{-1}})}
\end{equation}

The \citet{Wu05} work is based on a global correlation without
separating the low metallicity sources from metal rich galaxies. As 
can be seen in Fig. 4, a good agreement exists between the radio and
IR estimated SFRs (indicated by diamonds). If we use the more recent
calibration by \citet{Calzetti07} on SFRs from 24\,$\mu$m
luminosities, we find that most of the mid-IR estimated SFRs (marked
as filled circles) are located below the 1:1 proportionality line, and
thus they are consistently lower than both the SFRs estimated from the
radio or from \citet{Wu05}.  This is not unexpected since eq. 2 is
calibrated based on the high metallicity sources that have significant
24\,$\mu$m emission, while most of our sources are metal-poor galaxies.
We should also note that \citet{Calzetti07} measure the SFR in
individual H{\sc ii} regions in apertures within disks and subtract a
``disk background'', which could partly explain the deviations we see
from our analysis.  These authors have also mentioned that the SFRs of
low metallicity galaxies would be underestimated by a factor of 2--4
depending on how metal-poor the galaxies are. This deviation from a
linear correlation is likely due to the lower opacities for decreasing
metal content \citep{Walter07}.  When we fit the filled circles on
Fig. 4, we find that log(SFR[IR])=0.94$\times$log(SFR[1.4GHz]/3.98), which is
consistent with the metallicity correction factor suggested by
\citet{Calzetti07}. We should stress once more though that as
\citet{Bell03} has noted, the good agreement between the IR and radio
estimates of SFRs does not necessarily mean that these are the
``true'' SFRs for the dwarf galaxies we study, but rather the
competition of the lower dust content and suppressed synchrotron
emission balances each other.

\begin{figure}
  \epsscale{1.2}
  \plotone{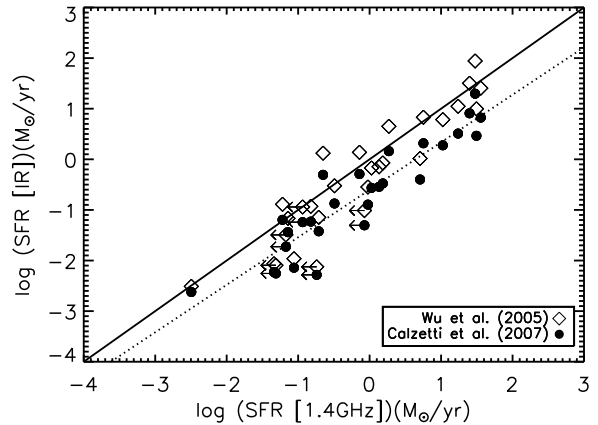}
  \caption{A plot of the SFR based on the 24\,$\mu$m luminosity as a
    function of the corresponding SFR estimated from the 1.4\,GHz
    radio continuum emission for each galaxy in our sample.  The open
    diamonds indicate the SFRs calculated using the 24\,$\mu$m fluxes
    following \citet{Wu05} (eq.1), the filled circles are those using
    eq.2 from \citet{Calzetti07}. The solid line is the 1:1
    proportionality line.  The dotted line is a fit to the filled
    circles.}
  \label{fig:fig4}
\end{figure}

\subsection{The Two Extremes: IZw18 and SBS0335-052E}

Despite this overall agreement between the IR/radio correlation in our
dwarf sample and the corresponding values of normal galaxies, the two
lowest metallicity galaxies in our sample, IZw18 and SBS0335-052E,
deviate markedly from the correlations and in opposite directions (see
Fig. 3\footnote{In Fig. 2 the ``outlier'' status of SBS0335-052E does
  not appear as prominent as in Fig. 3, because its high 24\,$\mu$m
  luminosity places it in a region of the plot where other sources
  with higher metallicity and higher luminosity are located.} ). As the
two most well-studied BCDs, it is interesting to inspect the IR/radio
properties of these two galaxies and a comparison of their physical
parameters can be found in Table \ref{tab3}.

\begin{deluxetable}{llcc}
\tabletypesize{\scriptsize}
\setlength{\tabcolsep}{0.2in}
\tablecaption{Comparison of SBS\,0335-052E and I\,Zw\,18\label{tab3}}
\tablewidth{0pc}
\tablehead{ 
  \colhead{}  & \colhead{SBS\,0335-052E} & \colhead{I\,Zw\,18} & \colhead{Ratio\tablenotemark{a}}\\
}
\startdata
12+log(O/H)            & 7.32    &   7.17   &  1.4  \\
D (Mpc)                & 58      &   18     &  3    \\
R (pc)                 & 560\tablenotemark{b} & 870\tablenotemark{c} & $\sim$0.6 \\
SFR(M$_\odot$\,yr$^{-1}$)& 7$\times$10$^{-1}$     &   5$\times$10$^{-2}$   &  14    \\
SNR(yr$^{-1}$)          & 6$\times$10$^{-3}$   &   2$\times$10${-4}$ &  30    \\ 
L$_{\rm H\alpha}$(L$_\odot$) & 5.6$\times$10$^7$\tablenotemark{d}  &  1.6$\times$10$^6$    &   35   \\
L$_{24\mu m}$(L$_\odot$)  & 9$\times$10$^8$   &   7$\times$10$^6$   & 120   \\
L$_{\rm IR}$(L$_\odot$)   & $\sim$4$\times$10$^9$   &  $\sim$4$\times$10$^7$  &  100   \\
L$_{\rm 1.4GHz}$(W\,Hz$^{-1}$)    & 2$\times$10$^{20}$   &  8$\times$10$^{19}$   & 2.5   \\
M(Ks)(mag)             & -18.3   &  -16.1   &  $\sim$80\tablenotemark{e}  \\

\enddata
\tablenotetext{a}{Ratio of the relative quantities in SBS0335-052E to IZw18.}  
\tablenotetext{b}{This is the size of the six super star clusters in SBS0335-052E, 
  which has a size of 2$\arcsec$ \citep{Thuan97}.  At a distance of 58\,Mpc, 
  1$\arcsec$ $\sim$280\,pc.}  
\tablenotetext{c}{This indicate the largest projected linear extend of the total 
  star-forming regions in this galaxy \citep[$\sim$10$\arcsec$,][]{Hunt05b}. At 
  a distance of 18\,Mpc, 1$\arcsec$ $\sim$ 87\,pc.}.
\tablenotetext{d}{Flux in 1$\arcsec$ slit multiplied by a factor of 2 to correct for aperture effects.}
\tablenotetext{e}{This is the approximate mass ratio of the two sources assuming 
  that their mass scales with K-band luminosity and that both sources have the 
  same M/L$_{\rm Ks}$ \citep[see][]{Bell01}.}

\end{deluxetable}

As indicated by the {\em Spitzer} IRS spectrum of SBS0335-052E
\citep{Houck04b}, the SED of the galaxy has an unusually flat mid-IR
slope that peaks at $\sim$28$\mu$m.  \citet{Hunt05a} used {\em DUSTY}
models \citep{Ivezic99} to fit its SED and found a q$_{\rm FIR}$ value
of 2.0 for this galaxy, lower than the typical q$_{\rm FIR}$ of
$\sim$2.3 expected for late type systems. These calculations are based
on a 1.4GHz continuum flux of $\sim$0.46\,mJy, which corresponds to
the compact part (6\arcsec) of SBS0335-052E \citep{Dale01, Hunt04}.
Similarly, if we simply calculate the q ratio independent of any
modelling work and only use the observable parameters, we find a
q$_{70}=$ 2.1$\pm$0.1 \citep[assuming that the MIPS 70\,$\mu$m flux
density is 51.1$\pm$4.8\,mJy,][]{Engelbracht07}. This is in good
agreement with the q$_{70}=$ 2.16$\pm$0.17 for the sources in the
First Look Survey studied by \citet{Appleton04}.

However, when we examine the q$_{24}$ for SBS0335-052E, adopting a
MIPS 24\,$\mu$m flux of 66\,mJy, we find a q$_{24}$ ratio of
2.2$\pm$0.1.  We see a 2$\sigma$ excess in the q parameter compared
with the mean q$_{24}$ of 1.3$\pm$0.4 in the dwarf sample. If this
excess is real, it would be consistent with the unique shape of the
SED of the galaxy, which suggests that the dust grain distribution is
dominated by small grains with temperatures of $\sim$150K and a total
dust mass of $\sim10^3$ M$_{\odot}$\citep{Houck04b}.  Whether the
large grains were never created or they were destroyed by shocks is
unknown.
The elevated q$_{24}$ ratio could also be due to the lack of
synchrotron emission. \citet{Roussel03,Roussel06} have shown their
study of three starburst galaxies, all of which have q$_{\rm FIR}$
more than 3$\sigma$ higher the mean q$_{\rm FIR}$ and categorized them
to be nascent starbursts. SBS0335-052E though, could also be a
candidate of this group. It is based partly on the high q$_{24}$
value, as well as its low H$\alpha$ luminosity as compared to the
infrared or radio luminosities (see Table \ref{tab3}).  Finally,
\citet{Hunt04} have shown the existence of free-free absorption in the
radio spectrum of SBS0335-052E, which is usually caused by young, 
dense and heavily embedded clusters\footnote{Note, that if we could
  properly account for the self-absorption of the radio continuum, the
  intrinsic q ratio would decrease. However, the exact fraction for
  absorption is not known and more data at $\nu<$1.5GHz would be
  needed to determine that number.}.

IZw18, despite its similar metallicity to SBS0335-052E
(12+log(O/H)=7.17 and 7.32 respectively), shows a rather different q
ratio. It also has an IR luminosity of L$_{\rm
  IR}$$\sim$10$^7$L$_\odot$, just 1\% of the L$_{\rm IR}$ in
SBS0335-052E.  In low luminosity galaxies, radio emission is known to
decrease faster than dust emission \citep{Devereux89}, and high
IR/radio ratios have been observed in low luminosity dwarf galaxies
\citep{Klein84}.
In IZw18 though, we find a rather low infrared to radio ratio. The
q$_{24}$ is found to be 0.5$\pm$0.1, nearly 2$\sigma$ lower than the
average value of our sample. At longer wavelengths, IZw18 is faint and
no {\em IRAS} FIR data are available.  However, as noted by
\citet{Wu07a} (see their Fig. 5), it has a very similar 5--38$\mu$m
continuum slope to the typical starburst galaxy NGC7714
\citep{Brandl04}.  If we were to assume that this similarity extends
to the FIR and scale down the {\em IRAS} 60 and 100\,$\mu$m flux densities
of NGC7714 by a factor of 375 so that its corresponding 22\,$\mu$m
flux density matches the one of IZw18, we find its ``equivalent''{\em
  IRAS} 60 and 100\,$\mu$m flux density to be 29.8 and 32.8\,mJy
respectively.  This would result in q$_{\rm FIR}$ of 1.3 for the
galaxy, which deviates from the average q$_{\rm FIR}$ ratio of
2.3$\pm$0.2 calibrated by \citet{Condon92} by nearly 5$\sigma$.  If we
were to use the MIPS 70\,$\mu$m detection of 34\,mJy for IZw18
\citep{Engelbracht07} and calculate its q ratio, we would find
q$_{70}$=1.4, again 5$\sigma$ away from the average q$_{70}$ suggested
by \citet{Appleton04}.

A number of plausible scenarios were considered to explain this
result, though none appears convincing. As discussed in \citet{Wu07a},
because the optical depth of IZw18 is small, it could be that a
significant fraction of the UV light has leaked out without being
absorbed by the dust, thus resulting in a damped 24\,$\mu$m emission.
Alternatively, IZw18 may simply have an unusually high radio
luminosity at 1.4\,GHz.  Could it be that IZw18 is observed at a
special moment right after the explosion of a supernovae event? Radio
supernovae fade by more than a factor of 10 within $\sim$3\,yr
\citep{Chevalier82}. The radio continuum observations of IZw18 span
over a period of more than 5\,yrs \citep{Hunt05b,Cannon05} but show no
variation, thus this is not likely.

The recent results by \citet{Murphy06b} provide another possible
scenario. These authors analyzed a sample of nearby spiral galaxies
and found that systems with higher disk-averaged SFRs ($\Sigma\,_{\rm
  SFR}$) have usually experienced a recent episode of enhanced star
formation. As a result they contain a higher fraction of young cosmic ray
electrons that have traveled only a few hundred parsecs from their
acceleration sites in supernova remnants.  This is perhaps the case
for IZw18. However, other sources in our sample also have elevated
$\Sigma\,_{\rm SFR}$ and show no radio excess. It could also be that
the extremely low metal abundance of IZw18, 0.15\,dex lower than that
of SBS0335-052E, is below a critical threshold. The newly formed
stars in the unpolluted interstellar medium may produce, and
subsequently heat, less dust than electrons which are accelerated and
contribute in the radio emission. We also note that the rate of SNR in
IZw18 is a factor of $\sim$30 lower than that in SBS0335-052E (see
Table \ref{tab3}) but its morphology is much more disturbed with
filamentary structure and outflows \citep[see][]{Izotov04}. If we were
to assume that both sources have similar mass-to-light ratios and
estimate their mass from their K-band luminosities, we find that the
mass of IZw18 is almost a factor of 80 less than that of SBS0335-052E.
Thus the rate of SNR normalized with mass in IZw18 would be $\sim$2.5
times that of SBS0335-052E. Could it be that the disturbed morphology
of IZw18 in addition to its low metal content that it created the
conditions in the interstellar medium for this abnormally high radio
flux? The question remains open.

\section{Conclusions}

We have studied the mid-IR and FIR to radio correlation in a sample of
dwarf star-forming galaxies spanning a metallicity range from
7.2$<$12+log(O/H)$<$8.9, using {\em Spitzer} IRS/MIPS, as well as radio
1.4\,GHz data obtained from the literature.  The BCD sample appears to
follow the same FIR/radio correlation as normal star forming galaxies.
The analysis based on mid-IR and radio data reveals a similar
correlation, though the scatter is larger, probably due to an
intrinsically higher variation in the 15-30$\mu$m SEDs of dwarf
galaxies.  When comparing the q$_{24}$ ratios with metallicity or
effective dust temperature, we find no strong correlation, though there
is a general trend of lower q ratios at lower metallicity for
galaxies with 12+log(O/H)$<$8.0 and the correlation flattens out
toward higher metallicity. In general the SFRs estimated from the
radio 1.4\,GHz continuum and the mid-IR data are in good agreement.
Two extremely metal poor BCDs, IZw18 and SBS0335-052E appear to
deviate from the average q$_{24}$ by $\sim$2$\sigma$, with one
displaying a radio excess and the other an infrared excess.

\acknowledgments 

The authors would like to thank P. Appleton and D. Calzetti for
stimulating discussions. We would also like to thank G. Helou, L.K.
Hunt as well as an anonymous referee whose detailed comments and
insightful suggestions have helped to improve this manuscript. This
work is based in part on observations made with the {\em Spitzer} Space
Telescope, which is operated by the Jet Propulsion Laboratory,
California Institute of Technology, under NASA contract 1407. Support
for this work was provided by NASA through Contract Number 1257184
issued by JPL/Caltech. VC would like to acknowledge the partial
support from the EU ToK grant 39965.


\begin{thebibliography}{}
\bibitem[Allende Prieto et al.(2001)]{Prieto01} 
  Allende Prieto, C., Lambert, D.~L., \& Asplund, M.\ 2001, \apjl, 556, L63 

\bibitem[Aloisi et al.(2007)]{Aloisi07}
  Aloisi, A. et al.\ 2007, \apjl, 667, L151

\bibitem[Appleton et al.(2004)]{Appleton04} 
  Appleton, P.~N., et al.\ 2004, \apjs, 154, 147 

\bibitem[Becker et al.(1995)]{Becker95} 
  Becker, R.~H., White, R.~L., \& Helfand, D.~J.\ 1995, \apj, 450, 559 

\bibitem[Bell \& de Jong(2001)]{Bell01} 
  Bell, E.~F., \& de Jong, R.~S.\ 2001, \apj, 550, 212

\bibitem[Bell(2003)]{Bell03} 
  Bell, E.~F.\ 2003, \apj, 586, 794 

\bibitem[Bergvall \& \"{O}stlin(2002)]{Bergvall02} 
  Bergvall, N., \& \"{O}stlin, G.\ 2002, \aap, 390, 891 

\bibitem[Brandl et al.(2004)]{Brandl04} 
  Brandl, B.~R., et al.\ 2004, \apjs, 154, 188 

\bibitem[Brandl et al.(2006)]{Brandl06} 
  Brandl, B.~R., et al.\ 2006, \apj, 653, 1129 

\bibitem[Boyle et al.(2007)]{Boyle07} 
  Boyle, B.~J., Cornwell, T.~J., Middelberg, E., Norris, R.~P., Appleton, P.~N., \& Smail, I.\ 2007, \mnras, 376, 1182 

\bibitem[Calzetti et al.(2007)]{Calzetti07} 
  Calzetti, D., et al. \apj, 666,870

\bibitem[Cannon et al.(2005)]{Cannon05} 
  Cannon, J.~M.,  Walter, F., Skillman, E.~D., \& van Zee, L.\ 2005, \apjl, 621, L21 

\bibitem[Chevalier(1982)]{Chevalier82} 
  Chevalier, R.~A.\ 1982, \apj, 258, 790 

\bibitem[Condon et al.(1991)]{Condon91} 
  Condon, J.~J., Anderson, M.~L., \& Helou, G.\ 1991, \apj, 376, 95 

\bibitem[Condon(1992)]{Condon92} 
Condon, J.~J.\ 1992, \araa, 30, 575 

\bibitem[Condon et al.(1998)]{Condon98} Condon, J.~J., Cotton, W.~D.,
  Greisen, E.~W., Yin, Q.~F., Perley, R.~A., Taylor, G.~B., \&
  Broderick, J.~J.\ 1998, \aj, 115, 1693

\bibitem[Dale et al.(2001)]{Dale01} 
  Dale, D.~A., Helou, G., Neugebauer, G., Soifer, B.~T., Frayer, D.~T., \& Condon, J.~J.\ 2001, \aj, 122, 1736 

\bibitem[Dale et al.(2006)]{Dale06} 
  Dale, D.~A., et al.\ 2006, \apj, 646, 161 

\bibitem[de Jong et al.(1985)]{Dejong85} 
  de Jong, T., Klein, U., Wielebinski, R., \& Wunderlich, E.\ 1985, \aap, 147, L6 

\bibitem[Devereux \& Eales(1989)]{Devereux89} 
  Devereux, N.~A., \& Eales, S.~A.\ 1989, \apj, 340, 708 


\bibitem[Elbaz et al.(2002)]{Elbaz02} 
  Elbaz, D., Cesarsky, C.~J., Chanial, P., Aussel, H., Franceschini, A., Fadda, D., \& Chary, R.~R.\ 2002, \aap, 384, 848 

\bibitem[Engelbracht et al.(2005)]{Engelbracht05} 
  Engelbracht, C.~W., Gordon, K.~D., Rieke, G.~H., Werner, M.~W., Dale, D.~A., \& Latter, W.~B.\ 2005, \apjl, 628, L29 

\bibitem[Engelbracht et al.(2007)]{Engelbracht07}
  Engelbracht, C.~W., et al.\ 2007, \apj, submitted

\bibitem[Garrett(2002)]{Garrett02} 
  Garrett, M.~A.\ 2002, \aap, 384, L19

\bibitem[Gruppioni et al.(2003)]{Gruppioni03} 
  Gruppioni, C., Pozzi, F., Zamorani, G., Ciliegi, P., Lari, C., Calabrese, E., La Franca, F., \& Matute, I.\ 2003, \mnras, 341, L1

\bibitem[Guseva et al.(2000)]{Guseva00} 
  Guseva, N.~G., Izotov,    Y.~I., \& Thuan, T.~X.\ 2000, \apj, 531, 776 

\bibitem[Hao et al.(2007)]{Hao07} 
  Hao, Lei, et al.,\ 2007, in  preparation

\bibitem[Harwit \& Pacini(1975)]{Harwit75} 
  Harwit, M., \& Pacini, F.\ 1975, \apjl, 200, L127 

\bibitem[Heckman et al.(1998)]{Heckman98} 
  Heckman, T.~M., Robert, C., Leitherer, C., Garnett, D.~R., \& van der Rydt, F.\ 1998, \apj, 503, 646 

\bibitem[Helou et al.(1985)]{Helou85} 
  Helou, G., Soifer, B.~T., \& Rowan-Robinson, M.\ 1985, \apjl, 298, L7

\bibitem[Helou \& Bicay(1993)]{Helou93} 
  Helou, G., \& Bicay, M.~D.\ 1993, \apj, 415, 93

\bibitem[Hopkins et al.(2002)]{Hopkins02} 
  Hopkins, A.~M., Schulte-Ladbeck, R.~E., \& Drozdovsky, I.~O.\ 2002, \aj, 124, 862 

\bibitem[Houck et al.(2004a)]{Houck04a} 
  Houck, J.~R., et al.\ 2004a, \apjs, 154, 18 

\bibitem[Houck et al.(2004b)]{Houck04b} 
  Houck, J.~R., et al.\ 2004b, \apjs, 154, 211 

\bibitem[Hummel et al.(1988)]{Hummel88} 
  Hummel, E., Davies, R.~D., Pedlar, A., Wolstencroft, R.~D., \& van der Hulst, J.~M.\ 1988, \aap, 199, 91 

\bibitem[Hunt et al.(2004)]{Hunt04} 
  Hunt, L.~K., Dyer, K.~K., Thuan, T.~X., \& Ulvestad, J.~S.\ 2004, \apj, 606, 853 

\bibitem[Hunt et al.(2005a)]{Hunt05a} 
  Hunt, L., Bianchi, S., \& Maiolino, R.\ 2005b, \aap, 434, 849 

\bibitem[Hunt et al.(2005b)]{Hunt05b} 
  Hunt, L.~K., Dyer, K.~K., \& Thuan, T.~X.\ 2005a, \aap, 436, 837 

\bibitem[Izotov \& Thuan(2004)]{Izotov04} 
  Izotov, Y.~I., \& Thuan, T.~X.\ 2004, \apj, 616, 768 


\bibitem[Izotov \& Thuan(1998)]{Izotov98} 
  Izotov, Y.~I., \& Thuan, T.~X.\ 1998, \apj, 500, 188 

\bibitem[Izotov \& Thuan(1999)]{Izotov99} 
  Izotov, Y.~I., \& Thuan, T.~X.\ 1999, \apj, 511, 639 

\bibitem[Ivezi\'c et al.(1999)]{Ivezic99} 
  Ivezi\'c, Z., Nenkova, M., \& Elitzur,M. 1999, User Manual for {\em DUSTY}, 
  University of Kentucky Internal Report,accessible at {\tt http://www.pa.uky.edu/~moshe/dusty} 

\bibitem[Jannuzi \& Dey(1999)]{Jannuzi99} 
  Jannuzi, B.~T.,~\& Dey, A.\ 1999, ASP Conf.~Ser.~191:
  Photometric Redshifts and the Detection of High Redshift Galaxies, 111

\bibitem[Kennicutt(1998)]{Kennicutt98} 
  Kennicutt, R.~C., Jr.\ 1998, \araa, 36, 189 


\bibitem[Kirshner et al.(1981)]{Kirshner81} 
  Kirshner, R.~P., Oemler, A., Jr., Schechter, P.~L., \& Shectman, S.~A.\ 1981, \apjl, 248, L57 

\bibitem[Klein et al.(1991)]{Klein91} 
  Klein, U., Weiland, H., \& Brinks, E.\ 1991, \aap, 246, 323 

\bibitem[Klein et al.(1984)]{Klein84} 
  Klein, U., Wielebinski, R., \& Thuan, T.~X.\ 1984, \aap, 141, 241 

\bibitem[Kobulnicky \& Skillman(1997)]{Kobulnicky97} 
  Kobulnicky, H.~A., \& Skillman, E.~D.\ 1997, \apj, 489, 636 

\bibitem[Kobulnicky \& Johnson(1999)]{Kobulnicky99} 
  Kobulnicky, H.~A., \& Johnson, K.~E.\ 1999, \apj, 527, 154 

\bibitem[Kunth \& Joubert(1985)]{Kunth85} 
  Kunth, D., \& Joubert, M.\ 1985, \aap, 142, 411 

\bibitem[Leroy et al.(2005)]{Leroy05} 
  Leroy, A., Bolatto, A.~D., Simon, J.~D., \& Blitz, L.\ 2005, \apj, 625, 763 

\bibitem[Miller \& Owen(2001)]{Miller01} 
  Miller, N.~A., \& Owen, F.~N.\ 2001, \aj, 121, 1903 

\bibitem[Moshir \& et al.(1990)]{Moshir90} 
  Moshir, M., \& et al.\ 1990, IRAS Faint Source Catalogue, version 2.0 (1990)

\bibitem[Moustakas \& Kennicutt(2006)]{Moustakas06} 
  Moustakas, J., \& Kennicutt, R.~C., Jr.\ 2006, \apjs, 164, 81 

\bibitem[Murphy et al.(2006a)]{Murphy06a} 
  Murphy, E.~J., et  al.\ 2006a, \apj, 638, 157

\bibitem[Murphy et al.(2006b)]{Murphy06b} 
  Murphy, E.~J., et  al.\ 2006b, \apjl, 651, L111 

\bibitem[Popescu \& Hopp(2000)]{Popescu00} 
  Popescu, C.~C., \& Hopp, U.\ 2000, \aaps, 142, 247

\bibitem[Rieke et al.(2004)]{Rieke04} 
  Rieke, G.~H., et al.\ 2004, \apjs, 154, 25 

\bibitem[Rosenberg et al.(2006)]{Rosenberg06} 
  Rosenberg, J.~L., Ashby, M.~L.~N., Salzer, J.~J., \& Huang, J.-S.\
  2006, \apj, 636, 742

\bibitem[Rosenberg et al.(2007)]{Rosenberg07} Rosenberg, J.~L., Wu,
  Yanling, Le Floc'h, E., Charmandaris, V., Ashby, M.~L.~N., Houck,
  J.~R., Salzer, J.~J. Willner, S.P.\ 2008, \apj, in press, astro-ph/0710.5514

\bibitem[Roussel et al.(2003)]{Roussel03} 
  Roussel, H., Helou, G., Beck, R., Condon, J.~J., Bosma, A., Matthews, K., \& Jarrett, T.~H.\ 2003, \apj, 593, 733 

\bibitem[Roussel et al.(2006)]{Roussel06} 
  Roussel, H., et al.\ 2006, \apj, 646, 841 

\bibitem[Sanders et al.(2003)]{Sanders03} 
  Sanders, D.~B., Mazzarella, J.~M., Kim, D.-C., Surace, J.~A., \& Soifer, B.~T.\ 2003, \aj, 126, 1607 

\bibitem[Sanders et al.(2007)]{Sanders07} 
  Sanders, D.~B., et al.\ 2007, \apjs, (in press arXiv:astro-ph/0701318)

\bibitem[Shi et al.(2005)]{Shi05} 
  Shi, F., Kong, X., Li, C., \& Cheng, F.~Z.\ 2005, \aap, 437, 849 

\bibitem[Storchi-Bergmann et al.(1994)]{Storchi94} 
  Storchi-Bergmann, T., Calzetti, D., \& Kinney, A.~L.\ 1994, \apj, 429, 572 

\bibitem[Thuan \& Izotov(2005)]{Thuan05} 
  Thuan, T.~X., \& Izotov, Y.~I.\ 2005, \apjs, 161, 240 

\bibitem[Thuan et al.(1997)]{Thuan97} 
  Thuan, T.~X., Izotov, Y.~I., \& Lipovetsky, V.~A.\ 1997, \apj, 477, 661 

\bibitem[van der Kruit(1971)]{van71} 
  van der Kruit, P.~C.\ 1971, \aap, 15, 110 


\bibitem[Walter et al.(2007)]{Walter07} 
  Walter, F., et al.\ 2007, \apj, 661, 102 

\bibitem[Werner et al.(2004)]{Werner04} 
  Werner, M.~W., et al.\ 2004, \apjs, 154, 1 

\bibitem[Wu et al.(2005)]{Wu05} 
  Wu, H., Cao, C., Hao, C.-N., Liu, F.-S., Wang, J.-L., Xia, X.-Y., Deng, Z.-G., \& Young, C.~K.-S.\ 2005, \apjl, 632, L79 

\bibitem[Wu et al.(2006)]{Wu06} 
  Wu, Y., Charmandaris, V., Hao, L., Brandl, B.~R.,
  Bernard-Salas, J., Spoon, H.~W.~W., \& Houck, J.~R.\ 2006, \apj,
  639, 157

\bibitem[Wu et al.(2007a)]{Wu07a}
  Wu, Y., et al., 2007a, \apj, 662, 952

\bibitem[Wu et al.(2007b)]{Wu07b} 
  Wu, Y.,  Bernard-Salas, J., Charmandaris,V. Lebouteiller, V., Hao,
  L., Brandl, B.R., Houck, J.R. \ 2007b, ApJ in press, 
  astro-ph/0710.0003

\bibitem[Yun et al.(2001)]{Yun01} 
  Yun, M.~S., Reddy, N.~A., \& Condon, J.~J.\ 2001, \apj, 554, 803 

\end{thebibliography}
\end{document}